\documentclass[12pt]{spieman}  
\usepackage{amsmath,amsfonts,amssymb}
\usepackage{graphicx}
\usepackage{setspace}
\usepackage{tocloft}
\usepackage{lineno}

\usepackage{tikz}
\usepackage{pgfplots}
\usetikzlibrary{plotmarks}
\pgfplotsset{width=10cm,compat=1.17}

\usepackage{newtxtext,newtxmath, bm}
\usepackage{subfig}

\DeclareMathOperator*{\tr}{tr}

\def\bA{{\bf A}}

\def\R {{\mathbb R}}

\title{The power of prediction: spatiotemporal Gaussian process modeling for predictive control in slope-based wavefront sensing}

\author[a, b,*]{Jalo Nousiainen}
\author[a]{Juha-Pekka Puska}
\author[b]{Tapio Helin}
\author[a]{Nuutti Hyvönen}
\author[c]{Markus Kasper}

\affil[a]{Aalto University, Department of Mathematics and Systems Analysis, P.O. Box 11100, FI-00076 Aalto, Finland}
\affil[b]{LUT University, Yliopistonkatu 34, FI-53850, Lappeenranta, Finland}
\affil[c]{European Southern Observatory, Karl-Schwarzschild-Str. 2, 85748, Garching bei M\"{u}nchen, Germany}

\cftpagenumbersoff{figure}
\cftpagenumbersoff{table} 
\begin{document} 
\maketitle
\begin{abstract}
Time-delay error is a significant error source in adaptive optics (AO) systems. It arises from the latency between sensing the wavefront and applying the correction. Predictive control algorithms reduce the time-delay error, providing significant performance gains, especially for high-contrast imaging. However, the predictive controller's performance depends on factors such as the WFS type, the measurement noise level, the AO system's geometry, and the atmospheric conditions.

This work studies the limits of prediction under different imaging conditions through spatiotemporal Gaussian process models. The method provides a predictive reconstructor that is optimal in the least-squares sense, conditioned on the fixed times series of WFS data and our knowledge of the atmospheric conditions. We demonstrate that knowledge is power in predictive AO control. With an SH-sensor-based extreme adaptive optics instrument, perfect knowledge of the wind and atmospheric profile and exact FF evolution lead to a reduction of the residual wavefront phase variance up to a factor of 3.5 compared to a non-predictive approach. If there is uncertainty in the profile or evolution models, the gain is more modest. Still, assuming that only effective wind speed is available (without direction) led to reductions in variance by a factor of about 2.3.

We also study the value of data for predictive filters by computing the experimental utility for different scenarios to answer questions such as: How many past telemetry frames should the prediction filter consider, and is it always most advantageous to use the most recent data? We show that within the scenarios considered, more data provides a consistent increase in prediction accuracy. Further, we demonstrate that given a computational limitation on how many past frames we can use, an optimized selection of $n$ past frames leads to a 10-15\% additional improvement in RMS over using the $n$ latest consecutive frames of data.

\end{abstract}



{\noindent \footnotesize\textbf{*} Jalo Nousiainen,  \linkable{jalo.nousiainen@eso.org} }

\begin{spacing}{2}   



\section{Introduction}
On ground-based telescopes, atmospheric turbulence causes variations in the optical path length of the incoming light hampering the telescope's image quality. Adaptive optics (AO) is a technique used to compensate for these variations \cite{babcock1953possibility, roddier1999adaptive}. The basic principle is to use a star or multiple guide stars (artificial or natural) as a reference point to measure these variations with a wavefront sensor (WFS) and then use a deformable mirror (DM) to compensate for the variations. 

Different AO techniques have been developed for different astronomical observation scenarios. Some AO systems aim to produce a good correction on a wide field of view, while others provide good corrections in multiple directions simultaneously. This paper focuses on single conjugate AO (SCAO) that utilizes a single guide star to obtain an excellent correction on a narrow field of view close to the guide star.

Time-delay error, also known as SCAO servo-lag error, is a significant error source in adaptive optics systems. It arises from the overall latency between sensing the wavefront and applying the correction. The delay occurs for various reasons, such as the time required to process the raw data, integrate the wavefront sensor frame and calculate the control signal, as well as the DM response time. The impact of delay can be significant, especially for AO instruments dedicated to direct exoplanet imaging, such as the Gemini Planet Imager \cite{macintosh2014first} on the Gemini South telescope, SPHERE (Spectro-Polarimetric High-contrast Exoplanet REsearch  \cite{2019A&A...631A.155B}) instrument on the European Southern Observatory's Very Large Telescope, MagAO-X (Magellan Adaptive Optics eXtreme system \cite{males2018magao}), and SCExAO (Subaru Coronagraphic Extreme
Adaptive Optics \cite{jovanovic2015subaru}). On these instruments, the temporal delay of AO creates a halo of stellar light appearing on the science camera, hiding the much fainter (compared to the host star) exoplanet beneath it. This phenomenon, known as the wind-driven halo (WDH), occurs particularly in windy conditions when atmospheric turbulence is in rapid motion. \cite{cantalloube2019peering}.

Time-delay error can be reduced with predictive control algorithms that use past telemetry data to predict incident phase aberrations at the time when the correction is applied to the DM. It is well-known that under strong assumptions on Markovian dynamics and linear system response, linear quadratic (LQ) control and Kalman filtering-based prediction yield asymptotically optimal predictive control \cite{kulcsar2006optimal}. However, the Markovian dynamics emerging from the Frozen Flow (FF) hypothesis with a single- or multi-layered turbulence offer only an approximate model (near Markovian) in the discretized state space.



More recently, data-driven predictive methods that use a longer time series of so-called \emph{pseudo-open loop telemetry} data have been proposed in the literature and show promising results in numerical simulations, optical bench, or on-sky settings (see, for example, Refs. \citenum{lloyd1996spatio, guyon2017adaptive, van2022predictive}). These approaches separate the reconstruction and prediction steps, where the prediction can be made either on WFS data or reconstructed DM commands.  As the data-driven methods often rely on longer time series of telemetry data than the auto-regressive models used with LQG controllers, it begs the questions: What are the limits of prediction beyond Markovian dynamics assumptions? What are the fundamental limits of such predictive control methods in AO, and how do these connect to other errors in wavefront estimation, such as spatial aliasing?

\begin{figure}
\centering
\includegraphics[trim={5cm 10cm 5cm 10cm}, clip, width=0.5\textwidth]{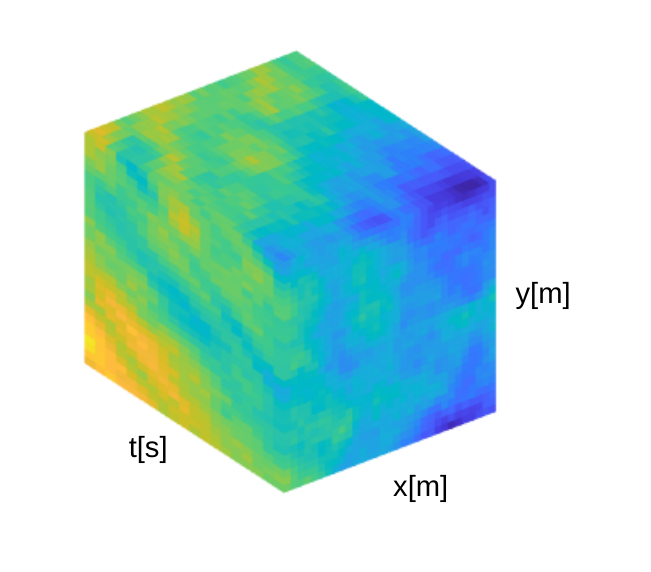}%
\caption{A stack of spatial turbulence fields at consecutive time steps. When expanded to the temporal axis, the cumulative spatial phase aberrations can be modeled as a spatiotemporal Gaussian process. The resulting process is stationary and non-isotropic, with a covariance function that depends on spatial atmospheric parameters (fried parameter, $C_N^2$ profile, $L_0$) and temporal parameters (wind speed and direction).} 
\label{fig:concept}
\end{figure}

We approach these questions by studying predictive controllers from the perspective of regression analysis beyond Markovian state space models. Specifically, we couple the wavefront reconstruction and prediction steps under a single prediction problem to achieve theoretical performance limits for predictive controllers with different levels of assumptions on the turbulence. We utilize Gaussian process (GP) regression \cite{rasmussen2006gaussian} to obtain principled uncertainty quantification for the predictive estimates given a time series of past observations. GP regression, also known as kriging in spatial statistics, is a technique used for regression tasks with complex relationships between variables. It is a tempting paradigm for AO as von K{\`a}rm{\`a}n turbulence models are Gaussian processes defined by their spatiotemporal covariance functions or power spectral density, and versatile \emph{a priori} information regarding the turbulence flow statistics can be introduced in the inference task. As an example, we consider a spatiotemporal Gaussian process (see Figure \ref{fig:concept}) emerging from a multilayer FF turbulence. Such a probability distribution can be easily improved by hierarchical modeling to consider the uncertainty in the estimates concerning wind speeds and the $ C_N^2$ profile. In practice, the turbulence parameters can also be identified with external algorithms and/or devices \cite{morujao2023integrated, lehtonen2019real, ono2017statistics, neichel2014towards} such as stereo-SCIDAR \cite{osborn2016turbulence, laidlaw2020automated}.

This paper explores the limits of predictive accuracy in GP regression by introducing two GP prior distributions for the spatiotemporal turbulence process that capture distinct levels of information: The first (very optimistic) prior distribution uses a multilayer FF turbulence model with perfect knowledge of the dynamics (wind directions, speeds, $r_0$s of all layers). In contrast, the second more conservative, prior distribution represents a scenario where only isotropic temporal correlation can be modeled, i.e., only the average wind speed is known, with no information on wind directions and $C_N^2$ profile. We also show how the spatiotemporal correlations enable the reconstruction of frequencies above the WFS sampling frequency, lowering the aliasing error of the DM correction. The sensitivity to high-order frequencies depends on the discretization of the WFS model; we study these discretization errors propagated to the predictive control.

Finally, we study how the number of past WFS measurements used for the prediction and reconstruction affects the prediction in different noise and atmospheric conditions. We compare different settings by computing the Bayes cost associated with different time series lengths and parametrizations of the system. As expected, when more precise prior knowledge of the dynamics is available, prediction can benefit from longer telemetry time series. Finally, we demonstrate that given a computational limitation on how many past frames we can use, it is better to use a sparse temporal sampling than a series of the last consecutive frames.



\section{Related work}
This paper continues the development of predictive methods for AO control. Previous literature on the topic (see, for example, \citenum{fowler2023tempestas}) demonstrates the potential for improved performance and increased stability of AO systems when predictive control techniques are used. However, further research is needed to optimize these methods for specific applications, improve their accuracy and efficiency, and analyze their effectiveness under different atmospheric conditions.

Predictive control algorithms include various types of model-based control methods such as Kalman filtering-based linear-quadratic-Gaussian controllers (LQG, e.g., Refs. \citenum{kulcsar2006optimal, paschall1993linear, gray2012ensemble, conan1a2011integral, correia2010adapting,correia2010optimal, correia2017modeling, sinquin2020sky, massioni2011fast}), as well as, closely related optimal controllers like the $\mathcal{H}_2$ controller \cite{hinnen2007exploiting, hinnen2008data}. In these works, the atmospheric turbulence is usually modeled with an autoregressive model in a predefined state space (either modal or zonal), and the derivation of the model sometimes includes parameter identification from past telemetry. The optimal control strategies also allow modeling dynamics beyond atmospheric turbulence, such as DM dynamics. These concepts have also been extended to tomographic systems; see, for example, \citenum{petit2009linear, le2004optimal}.

More recently, neural-network-based reinforcement learning (RL) methods have been studied in various works \cite{nousiainen2021adaptive, nousiainen2022toward, nousiainen2024laboratory, landman2021self, Pou:22, pou2024integrating}. In RL, the control/system model is usually parametrized with a neural network, and the desired control is learned from the telemetry data with minimal prior assumptions on the system. However, RL is not limited to NN models, and similar strategies can be used with linear models, which leads to LQG-like derivation of the control law \cite{haffert2021data,haffert2021data1}. A mathematical analysis of the overlap between RL and optimal control strategies is available in \citenum{recht2019tour}.

Another approach is to separate the reconstruction and prediction steps of AO control, that is, use a predictive filter that operates on an open-loop estimate (pseudo-open-loop measurement/reconstruction on the closed-loop system) of the full turbulence \cite{guyon2017adaptive}. These predictive filters aim to forecast the future wavefront based on past wavefront measurements utilizing a variety of techniques, such as Fourier analysis and Zernike polynomials, as well as machine learning.  Fourier-based methods decompose the wavefront into its constituent frequency components and predict the future behavior of each component, while Zernike polynomials are a set of orthogonal functions that can be used to represent the wavefront and predict its future behavior (e.g., Refs. \citenum{poyneer2007fourier, males2018ground, dessenne1998optimization}). Machine learning methods can learn a predictive model of the wavefront based on historical data and make predictions based on the learned model (e.g., Refs. \citenum{guyon2017adaptive, swanson2018wavefront, swanson2021closed, sun2017bayesian, mcguire1999adaptive}).

This work is closely connected to minimum variance predictive controllers, where the prediction model is constructed by utilizing the temporal or spatiotemporal covariance structure of turbulence \cite{doelman2020minimum, correia2015spatio, lloyd1996spatio}. The cross-correlations can either be derived from the atmospheric model \cite{doelman2020minimum, correia2015spatio, lloyd1996spatio},  or by utilizing machine learning techniques \cite{guyon2017adaptive, van2019impact, van2022predictive}. On one hand, we expand the work of Doelman (2020)\cite{doelman2020minimum} from the temporal to the spatio-temporal domain, include the WFS model, and give estimates for optimal predictive control limit in all spatial locations on the telescope pupil. On the other hand, we discuss the spatio-temporal minimum variance prediction matrix (e.g., in \citenum{correia2015spatio}) beyond deriving the minimum variance predictor for a fixed Markov state space. In particular, we derive the theoretical limits of minimum variance predictive control, as well as study the effect of WFS modeling accuracy and the effect of the chosen past measurements on prediction accuracy under different imaging conditions and levels of prior information on the atmosphere.

Since, we also take into account that the turbulence itself is not observed directly but through a WFS measurement, which connects our considerations to works that study priors from the perspective of wavefront reconstruction (see, e.g., Ref. \citenum{bechet2009comparison}).
Moreover, methods that enable reconstruction above the sampling frequency of WFS have lately gained attention in the AO literature. Oberti et al. (2022) \cite{oberti2022super} discussed how data from several misaligned WFS could be used to obtain super-resolution reconstructions, and Berdeu et al. (2022) \cite{berdeu2022inverse} studied how the discretization of the reconstruction grid affects the aliasing error. The super-resolution has also been studied in the context of Pyramid WFS. Correia et al. (2022)\cite{correia2022super} showed how a single Pyramid WFS, with facets (the four pupil images on the detector plane) that are adjusted to offset, also enables super-resolution. Contrarily, we use single slope-based WFS and utilize the FF structure of turbulence and past telemetry to gain information on the high-order spatial frequencies.


\section{Atmospheric turbulence as a spatiotemporal Gaussian process}

\subsection{Spatial modeling}
The atmosphere is the main cause of distortion in the phase of the light that reaches ground-based telescopes. This distortion is caused by the random mixing of air at different temperatures, that is,  the atmospheric turbulence, constantly moving due to wind. This movement causes variations in the refractive index of the air and, thus, in the optical path length of the incoming light. The turbulence is typically modeled as a series of thin, independent layers at different heights with different levels of turbulence strength. As the light propagates through such a thin layer, the resulting phase variations $\phi(x,y)$ can be modeled by a two-dimensional homogeneous Gaussian process described by the von K{\`a}rm{\`a}n power spectral density (PSD)
\begin{equation}\label{eq:PSD}
\hat{k}_{\phi}(\bm \kappa) \propto \big(\vert \bm \kappa \vert^2 + 1/L_0^2\big)^{-\frac{11}{6}}.
\end{equation}
Here $\bm \kappa$ is the spatial frequency, and $L_0$ is the so-called outer scale of the layer. The PSD expresses the amount of turbulent energy at a given spatial frequency. By the Wiener–Khinchin theorem, the corresponding covariance function  $k_{\phi}(r)$ is given through the Fourier transform of the PSD. By computing the Fourier transformation, we get
\begin{equation}\label{coc}
k_{\phi}(r) = \frac{\Gamma\left(\frac{11}{6}\right)}{\sqrt{2}\pi^{\frac{8}{3}}}\left(\frac{24}{5}\Gamma\left(\frac{6}{5}\right)\right)^{\frac{5}{6}} \left( \frac{L_0}{r_0}\right)^{\frac{5}{3}}\left(\frac{2\pi r}{L_0}\right)^{\frac{5}{6}} K_{\frac{5}{6}}\left(\frac{2\pi r}{L_0}\right),
\end{equation}
where $r_0$ is the Fried parameter (defining the total turbulence strength), $\Gamma$ is the gamma function, and $K_{\frac{5}{6}}$ is the modified Bessel function of the second kind. For a detailed derivation of the PSD and the covariance function, see, e.g., Conan (2008)\cite{conan2008mean}. The final cumulative optical path aberrations $\Phi$ is the sum of the aberrations along the line of sight
\begin{equation}
\Phi(x,y) = \sum_{\ell=1}^L \sqrt{\rho_\ell} \phi_\ell(x,y),
\end{equation} 
where $L$ is the total number of layers, and $\rho_\ell$ and $\phi_\ell$ are, respectively, the relative strength and the phase aberrations of the $\ell$th turbulence layer. The collection of the relative strengths $[\rho_1, \dots, \rho_L]$ at the considered layers is called the discrete $C_N^2$ profile. The spatial turbulence statistics can be estimated, for example, from telemetry data \cite{morujao2023integrated}.

In what follows, we restrict our attention to the von K{\`a}rm{\`a}n model given in \eqref{eq:PSD}, but let us in any case remark that deviations from the von K{\`a}rm{\`a}n power law close to the ground \cite{bester1992atmospheric, dayton1992atmospheric} and in the upper troposphere and stratosphere \cite{belen1999experimental, kyrazis1994measurement, stribling1995optical} are well-documented. 
Similarly, the turbulence taking place in the telescope dome (so-called dome seeing) has different characteristics and is naturally dependent on the dome geometry \cite{lombardi2010surface}. While there are methods to estimate these deviations, such as proposed in Helin et al. (2018)\cite{helin2018atmospheric}, they are based on solving an ill-posed problem over a relatively long time series of telemetry data leading to inherent uncertainty about temporal fluctuations. The non-stationarity of the turbulence is discussed more thoroughly in \citenum{tokovinin2023elusive}.

\subsection{Spatiotemporal model}
\label{sec:stmodel}

On the millisecond scale of AO control, the changes in the turbulence pattern itself are small, and the temporal evolution is mainly driven by advection, i.e., wind at the altitudes of the layers. Hence, Taylor's FF hypothesis provides a good approximation to the time evolution; i.e., each turbulent layer is modeled as a thin static 'frozen' layer sliding over the telescope with an individual wind speed and direction. 

There have been attempts to involve also other physical effects from the underpinning Navier--Stokes equation, such as kinematic diffusion \cite{vogel2006time} and intermittency \cite{yuan2009time}. To our knowledge, these dynamics models have not been studied in a predictive control context.

We now formulate two spatiotemporal distributions that will be considered in the regression below. The first distribution is based on exact knowledge of the FF model, including the correct $ C_N^2$ profile, and it is therefore optimistic. In the second distribution, we assume that we have an estimate of the weight-averaged wind velocity over the atmosphere and, more importantly, that the spatial and temporal statistics are essentially treated independently since there is no deterministic flow originating from the spatial process.

We note that the first distribution is overly optimistic since it assumes perfect detection of FF model hyper-parameters. Poyneer et al. 2009 \cite{{poyneer2009experimental}} showed that even though FF was detected most of the time from Altair and Keck AO system telemetry, it only covered 20\% -- 40\% of the total controllable phase and originated
usually from one to three layers. The latter distribution is more realistic as it neglects strong spatiotemporal correlations emerging in advection-dominated time scales. This model still assumes stable coherence time and von K{\`a}rm{\`a}n spectrum, making it somewhat over-optimistic. Uncertainty in these hyperparameters can be added to the models similarly to uncertainty in the wind directions. The models can also be combined to simulate/predict where part of the FF cannot be identified or turbulence is only partially in FF but still follows stable coherence time.

\subsubsection{Frozen flow induced spatiotemporal Gaussian process}
Let us first introduce the GP model based on FF with the exact knowledge of the layered model. We will refer to it as the 
FF-GP for convenience.

Consider FF on a single layer. At a given initial time instance, $t = 0$, the phase aberrations $\psi(x,y, 0) = \phi(x,y)$ follow a two-dimensional GP statistics, i.e., $\phi(x,y) \sim GP(0,k_{\phi}(r))$. We expand the spatial GP to the time domain according to the FF model
\begin{equation}
\psi(x,y,t) = \phi(x + v_x t, y + v_y t),    
\end{equation}
where $\bm v = (v_x, v_y)$ stands for the two-dimensional wind velocity vector.
Consequently, the covariance function of the spatiotemporal process is obtained by
\begin{align*}
&k_{\psi}\big((x_1,y_1,t_1 ), (x_2,y_2,t_2)\big) \\&= k_\phi \left(\sqrt{(x_1 - x_2 + (t_1 - t_2)v_x)^2 - (y_1 - y_2 + (t_1 - t_2)v_y)^2}\right) .  
\end{align*}
The full multi-layer phase aberrations at the entrance pupil $\Psi$ are obtained by adding the single-layer spatiotemporal GPs together, that is,
\begin{equation}\label{eq:stgp}
\Psi_{\rm FF} \sim GP \left( 0, k_{\Psi_{\rm FF}}((x_1,y_1,t_1 ), (x_2,y_2,t_2))\right),\\
\end{equation}
where
\begin{align*}
    k_{\Psi_{\rm FF}}\big((x_1,y_1,t_1 ), & (x_2,y_2,t_2)\big)  \\
    = & \sum_{\ell=1}^L \rho_\ell k_{\psi_\ell}((x_1,y_1,t_1 ), (x_2,y_2,t_2)).
\end{align*}


\subsubsection{Wind-averaged frozen flow Gaussian process}

The information about atmospheric profile, including wind directions, can contain various uncertainties. Here, we assume that reliable information is only available regarding the weight averaged wind velocity $|\bm v_{\rm avg}|$ through the atmosphere.
In practise, reliable information is often available regarding $|\bm v_{\rm avg}|$ as, e.g., the coherence time $\tau_0$ is directly related to this quantity via
\begin{equation}
\tau_0 = 6.88^{-\frac 35} \frac{r_0}{|\bm v_{\rm avg}|},
\end{equation}    
where $r_0$ is the Fried parameter. The uncertainty regarding wind direction 
can be rephrased as an assumption that each direction is equally probable, i.e.,
the wind direction is uniformly distributed
$\bm \theta := {\bm v_{\rm avg}}/|\bm v_{\rm avg}| \sim U({\mathbb S}^1)$, where ${\mathbb S}^1$ stands for the unit circle. We will refer to this model as wind-averaged FF GP (WAFF-GP).

We formulate WAFF-GP as a zero-mean process $\Psi_{\rm WAFF} \sim GP(0,k_{\Psi_{\rm WAFF}}((x_1,y_1,t_1 ), (x_2,y_2,t_2))$, where
\begin{align}
\label{eq:wind-avg_covariance}
&k_{\Psi_{\rm WAFF}}\big((x_1,y_1,t_1 ), (x_2,y_2,t_2)\big) \\&
=\frac{1}{|{\mathbb S}^1|}\int_{{\mathbb S}^1} k_\phi \left(\sqrt{(x_1 - x_2 + \Delta  s\,\theta_x)^2 - (y_1 - y_2 +  \Delta s\, \theta_y)^2}\right) d\bm \theta, \nonumber
\end{align}
with the convention $\Delta s = (t_1-t_2)|\bm v_{\rm avg}|$ and $\bm \theta = (\theta_x, \theta_y) \in {\mathbb S}^1$. The covariance function specified by identity \eqref{eq:wind-avg_covariance} is well-defined as it is an average of 
well-defined covariances \cite{rasmussen2006gaussian}.

Note that the process formulated this way coincides with the spatial von K{\`a}rm{\`a}n statistics for any fixed time $t_1=t_2$, i.e.,
\begin{equation*}
    k_{\Psi_{\rm WAFF}}\big((x_1,y_1,t ), (x_2,y_2,t)\big) = k_\phi\big((x_1,y_1),(x_2,y_2)\big).
\end{equation*}
For the purpose of this study, WAFF-GP presents a more pessimistic model of spatiotemporal statistics, where less information is included than is often practically available. That being said, one could relax definition \eqref{eq:wind-avg_covariance} further by assuming a weighted average over $|\bm v_{\rm avg}|$ with respect to some probability density, e.g., a uniform distribution over a confidence interval $[|\bm v_{\rm avg}|-\epsilon, |\bm v_{\rm avg}|+\epsilon]$ modeling the measurement accuracy. Further, WAFF, and FF models can also be combined together to create a prior for conditions where we, for example, detect three FF layers that cover 40\% of the turbulence, and movement of the other layers cannot be detected.\cite{poyneer2009experimental}

\section{Reconstruction and prediction}


In this section, we describe the observational model in AO and how the Bayesian inference and prediction are carried out.

\subsection{AO system and wavefront sensing}\label{sec:wfs}

Single-conjugate AO systems use a single natural guide star as the reference source. The wavefront control comprises two main components: the WFS and the DM. The WFS measures the distortion caused by the atmosphere in the incoming phase along the line of sight, and the DM then compensates for these distortions by taking a shape that cancels them out. The WFS is usually set downstream from the DM, effectively measuring the deviation from a flat wavefront (closed-loop residuals). However, if the DM influence function and the control delay in the system are known, the open-loop measurement can be recovered through the so-called pseudo-open loop scheme.

This paper focuses on the open-loop setup, i.e., we assume that the WFS observes the full phase error caused by the atmosphere. We model the DM with Gaussian influence functions.


Next, let us describe the non-direct observation model. We assume that WFS measures the gradient of phase aberrations $\phi$ averaged over a period $\Delta \tau$ set by the AO system's framerate on a given spatial sampling defined by, e.g., the number of lenslets in the Shack--Hartman wavefront sensor (SHS): 
\begin{equation}
\label{eq:SHS_map}
w_i = \frac{1}{|B_i|} \int\limits_t^{t + \Delta \tau} \iint\limits_{B_i} \nabla \phi(x,y,t)dxdydt \in \R^2 
\end{equation}
where  $B_i$ is the sub-aperture surface indexed by $i = 1,\dots,M$. The concatenation of $w$ at all possible locations is denoted by a measurement vector $\bm w \in \R^{2M}$. However, the methods discussed in the following apply to any linear or linearly approximated WFS, where the mathematical model can be written out as a matrix.

Further, suppose our computational resources allow utilizing the latest $p$ subsequent data vectors with recording intervals of $\Delta t$ in the prediction task. Let us denote these vectors by
\begin{equation}
\label{eq:timenotation}
	 \bm w^1, \bm w^2, \dots , \bm w^p \in \R^{2M},
\end{equation}
with the (loose) convention $\bm w^k = \bm w(k\Delta t)$ for any time-dependent entity.

\subsection{Bayesian inference and prediction}\label{sec:stgp_posterior}

\begin{figure}
\centering
	\includegraphics[width=0.5\columnwidth]{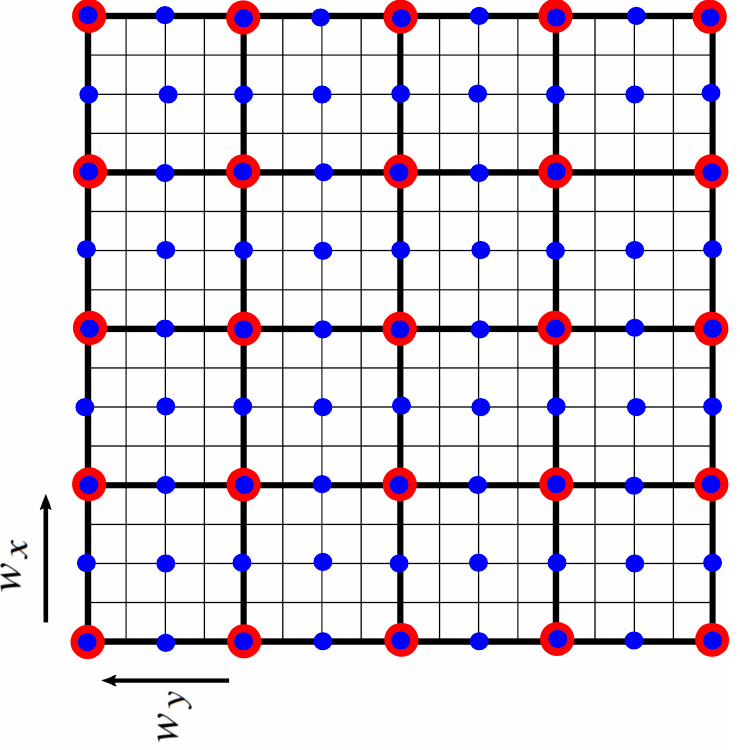}
    \vspace{0.2cm}
    \caption{This figure illustrates the different discretization of the state and measurement domains used in this study. Thick black lines depict a 4x4 SHS sensor, the red points represent the locations of the DM actuators in the Fried geometry, and the blue points are the so-called extended Fried grid where a set of pixels are added between Fried grid. The SHS measures the average gradient over each lenslet. This work considers discretization grids depicted by the thin black lines to various amounts of pixels ($4 \times 4$ (in the figure) to $8 \times 8$) and more coarse discretization grids defined by the Fried geometry and extended Fried Grid (red dots, blue dots).}
    \label{fig:fried}
\end{figure}

Let us consider the incoming phase aberrations and their representation on a discrete grid. We utilize evenly spaced spatial grids following the Fried geometry or a super-sampled grid where each SHS-lenslet consists of an evenly spaced pixel grid (e.g., $4 \times 4$ pixels or  $8 \times 8$ pixels  ); see Figure \ref{fig:fried}.

For any fixed time instance $t$, the state vector $\bm \phi (t)$ belongs to $\R^{N^2}$ (minus the inactive pixels). Here, $N^2$ is the number of pixels (either in the Fried or a super-resolution grid), that is,~$N$ is the number of spatial locations along a single spatial dimension. Let us write
\begin{equation*}
	\bm \phi^1, \bm \phi^2, \dots , \bm \phi^p \in \R^{N^2}
\end{equation*}
for the state vectors representing the phase on the different time steps, following the notation in Equation \eqref{eq:timenotation}.

Our aim is to predict/reconstruct $\bm \phi^{p+2}$ as the typical time lag in next-generation telescopes is around two-time steps (expanding the method to non-integer delays is trivial). Therefore, our state vector will include $\bm \phi^{p+2}$ as well.
Let us denote the concatenated state and data vectors as
\begin{equation*}
	\bm \Phi =
	\begin{pmatrix}
		\bm \phi^1 \\
		\vdots \\
		\bm \phi^{p} \\
        \bm \phi^{p+2}
	\end{pmatrix}
	\in \R^{(p+2)N^2} \quad \text{and} \quad
	\bm W = 
	\begin{pmatrix}
		\bm w^1 \\
		\vdots \\
		\bm w^p
	\end{pmatrix}
	\in \R^{2pM}.
\end{equation*}
We can formulate our prediction task as an inverse problem of solving for $\bm \Phi$ in \begin{equation}
    \label{eq:observational_model}
	\bm W = \bA \bm \Phi + \epsilon,
\end{equation}
where the forward operator is given as a block diagonal matrix
\begin{equation}
    \label{eq:forward_operator}
	\bA = 
	\begin{pmatrix}
		A & 0 & \hdots & 0 & 0  \\
		0 & A & &  \vdots & \vdots  \\
		\vdots  &   & \ddots & & \\
		0  & & & A & 0\\ 
	\end{pmatrix}
	\in \R^{pM \times (p+1)N},
\end{equation}
Each matrix $A$ in Equation \eqref{eq:forward_operator} maps an incoming state $\bm \phi^i$, $i=1,\dots ,p$, to the corresponding data vector $\bm w^i$ according to Equation \eqref{eq:SHS_map}. We model the noise $\epsilon$ as a zero-mean Gaussian random variable with a symmetric positive definite covariance matrix $C_{\rm noise}\in \R^{2pM \times 2pM}$.


In the Bayesian paradigm, the solution to an inference problem \eqref{eq:observational_model} is the conditional distribution of $\bm \Phi$ given observational data $\bm W$, i.e., the posterior distribution. By the Bayes' formula, the Gaussian prior models crafted in Section \ref{sec:stmodel} combined with the Gaussian likelihood yields a Gaussian posterior. 

Here, the Gaussian prior distribution over $\Phi$ has zero-mean and the covariance matrix $C_{\rm prior}$ given by 
\begin{equation*}
    (C_{\rm prior})_{ij} = k_{\Psi_{\rm FF/WAFF}}((x_i, y_i, t_i), (x_j, y_j, t_j)),    
\end{equation*}
where $i,j = 1,\dots,(p+2)N^2$ correspond to global indexing over the spatial and temporal variables.

It follows that the Gaussian posterior distribution is defined by the covariance matrix
\begin{equation}
\label{eq:stgp_posterior}
	C_{\rm post} = \left(\bA^\top C_{\rm noise}^{-1}\bA + C_{\rm prior}^{-1}\right)^{-1}
\end{equation}
and the mean vector
\begin{equation*}\label{eq:stgp_mean}
	\bm \Phi_{\rm post} = C_{\rm post} \bA^\top C_{\rm noise}^{-1} \bm W.
\end{equation*}
The sought-for prediction entails assessing the marginal posterior distribution of $\bm \phi^{p+2}$,~i.e.,~the last $N^2$ components in $\bm \Phi$. For convenience, let us denote by $P : \R^{(p+2)N^2} \to \R^{N^2}$ a matrix projection that maps the full state vector to the state at time $p+2$, i.e., $P \bm \Phi = \bm \phi^{p+2}$. It follows that the predictive posterior at time $p+2$ is given by a Gaussian distribution with the mean $P\bm \Phi_{\rm post}$ and the covariance matrix $PC_{\rm post}P^\top$.

Since the conditional mean and the MAP estimate coincide for a Gaussian posterior, the natural choice for the point estimate (i.e., the prediction to be used in control) is the mean value of the posterior. The spatiotemporal AO prediction matrix $R_{\rm FF/WAFF}$ thus takes the form
\begin{equation}\label{eq:recontructor}
R_{\rm FF/WAFF} =  P C_{\rm post} \bA^\top C_{\rm noise}^{-1}.
\end{equation}

Further, the corresponding DM commands are calculated as standard least-squares-fit to the DM influence function. Also, the posterior covariance $P\bm \Phi_{\rm post}$ can projected to DM space using the least-squares-fitting matrix.


\section{Bayesian utility of experimental designs}
Experimental design is the process of determining how to perform an experiment such that the informativeness of the data is maximized and, therefore, the uncertainty in the estimates produced is minimized. For a general review of Bayesian optimal experimental design (OED), we refer the reader to Chaloner \& Verdinelli (1995) \cite{chaloner1995bayesian}. For AO control, the design of the experiment can mean,~e.g.,~the choice of the length of telemetry time series that are given as input for the predictive controller as described below in Section \ref{ssec:oed}. It could also address other design parameters, though outside the scope of this paper, such as the number of sub-apertures to be included in the WFS or the modulation amplitude for a pyramid WFS in given conditions. 

In this paper, we study the informativeness of past WFS measurements on the predictive reconstruction quality. It is a naturally occurring question since the computational resources are scarce in AO, and one would naturally like to minimize the amount of data that is processed. Also, the GP model allows one to write the Bayesian expected loss/utility as a function of the posterior covariance matrix, as described below. These formulas are relatively efficient to evaluate and make computing optimal values for the decision variables feasible even for high-dimensional problems as encountered in AO.

Let us briefly review the basics of Bayesian OED in our framework. For this, assume that we have a parameter ${\bm d} \in \mathbb{R}^n$ that defines an experimental design for AO control, and let us account for the dependence of the forward model on the design parameter by writing the system matrix as $\bA({\bm d})$. One then chooses a loss function $u({\bm \Phi},{\bm W},{\bm d})$
and defines the expected loss (or negative utility) of the design as
\begin{equation*}
    U({\bm d}) = \mathbb{E}_{{\bm \Phi},{\bm W}}[u({\bm \Phi},{\bm W},{\bm d})],
\end{equation*}
for which small values indicate informative measurements (if the choice of the loss function is adequately chosen). Comparing different experimental designs involves computing $U({\bm d})$ (and possibly also its gradient) for a number of ${\bm d}$, which can be a nontrivial and computationally demanding task in general. However, in our linear and Gaussian setting, choosing the loss function as the quadratic loss
\begin{equation*}
    u({\bm \Phi},{\bm W},{\bm d}) = \| P({\bm \Phi}- {\bm \Phi}_{\rm post}({\bm W}, {\bm d})) \|_2^2,
\end{equation*}
results in the well-known A-optimality criterion. More precisely, the corresponding expected loss (or minimization target) reduces to 
\begin{equation}
    U({\bm d}) = \tr \big(P C_{\rm post}({\bm d})P^\top\big)
\end{equation}
where the dependence of $C_{\rm post}$ on ${\bm d}$ is inherited from $\bA({\bm d})$ via \eqref{eq:stgp_posterior}.
For derivations of these formulas, see, e.g., Burger et al. (2021)\cite{Burger21}.

In Section \ref{ssec:oed}, we will evaluate the utility of the data on the optimality criteria for the different prior models and different noise levels. Thus, the design variable ${\bm d}$ will be the choice of timesteps $p$ used for computing the posterior. Note that including more time steps in the prediction process always decreases the uncertainty about the unknown, thus reducing the expected error. However, if clearly diminishing return in the value of longer telemetry time series is observed, then one may conclude that the benefit from the usage of the extra data does not merit the required additional computational expense.

\section{Numerical experiments}
\label{sec:numerics}
This section demonstrates the performance of spatiotemporal GP prediction (FF-GP and WAFF-GP) through numerical experiments, where we utilize the HciPy toolbox \cite{por2018hcipy}. We design the numerical experiments to demonstrate three key features of the approach: the predictive capacity, the noise reduction (see Section \ref{sec:pred_and_noise}), and the ability to recover spatial frequencies above the sampling frequency of the WFS (see Section \ref{sec:antialias}). 

Further, we demonstrate how discretization of the measurement model affects the reconstruction quality and how the Bayes loss can be used to decide the right reconstructor design for the given conditions and AO system (see Section \ref{ssec:oed}). The quantity of interest in these experiments is mainly the variance of the posterior distribution that gives essentially the measure for the performance of different priors in different experiments. We compare FF-GP and WAFF-GP models to the standard minimum variance reconstruction that only considers spatial statistics, called spatial GP (S-GP). 

The three different reconstruction models are:
\begin{enumerate}
    \item \textbf{FF-GP:} the Frozen Flow GP prior that assumes perfect information about the FF and is defined by \eqref{eq:stgp}. 
    
    \item \textbf{WAFF-GP:} the Wind Averaged FF GP prior for unknown wind direction and $C_N^2$ profile defined by \eqref{eq:wind-avg_covariance}.
    
    \item \textbf{S-GP:} a spatial prior that only considers spatial information, that is, every time step is reconstructed separately. The model is non-predictive. 
\end{enumerate}
The Bayes formula straightforwardly gives the standard spatial reconstruction method S-GP for a single frame. It yields a Gaussian posterior distribution with the covariance matrix
\begin{equation}
\label{eq:sgp_posterior}
	C_{\rm post} = \left(A^\top C_{\rm noise}^{-1} A + C_{\rm spatial}^{-1}\right)^{-1}
\end{equation}
and the mean vector
\begin{equation}\label{eq:sgp_mean}
	\bm \phi_{\rm post}^{p} = C_{\rm post} A^\top \bm w^p.
\end{equation}
As mentioned above, this reconstruction technique does not consider temporal statistics, and it is, hence, always limited by the temporal delay of the AO system.

\subsection{Simulation set-up}

We simulated a 3.2-meter telescope with 14\% central obstruction and linear slope-based WFS (see Section \ref{sec:wfs}) with $16 \times 16$ lenslets (20cm actuator spacing). The WFS sampling is chosen to represent XAO systems.  Instead of simulating data by sliding two-dimensional FF layers and interpolating, we generate the data by constructing the spatiotemporal covariance matrix as explained in Section \ref{sec:stmodel} and sampling from the model. Generating data in this manner does not include any approximations, such as interpolating the temporal movement; that is, no high-order frequencies are dampened in the temporal movement.


The parameters for the von K{\`a}rm{\`a}n turbulence are $r_0 = 16.8$ cm and $L_0 = 20$ m. The time step between frames is set to $\Delta\tau = 2$~ms, and the atmosphere is composed of seven layers. A complete list of the parameter values employed in the simulations can be found in Tables~\ref{table:simulator_parameters2} and \ref{table:simulator_parameters3}.  

\begin{table}
    \centering
    \caption{ Simulation parameters}
    \label{table:simulator_parameters2}
\begin{tabular}{ |c| c| c|}
 \hline
 \multicolumn{3}{|c|}{Telescope} \\
 \hline
         Parameter  & Value  &  Units  \\
 \hline  
 Telescope diameter   &  3.2  & m     \\
 Obstruction ratio    &  14  & percent           \\
 Sampling frequency   &  500/1000Hz  & Hz        \\
 Measurement noise &   100/4   &     S/N     \%      \\ 
 WFS wavelength & 0.79 &  \textmu  m  \\
 WFS lenslets  & 16 & across the pupil \\
 Pixels & 128 & across the pupil \\
 DM actuators & 17 & across the pupil \\
 DM influence function & Gaussian & - \\
 DM coupling & 40 & percent \\
  \hline  
\end{tabular}
\end{table}

\begin{table}
    \centering
    \caption{Atmospheric parameters}
    \label{table:simulator_parameters3}
\begin{tabular}{ |c| c| c| c| c| }
 \hline
 \multicolumn{5}{|c|}{Atmosphere parameters ( 15 cm @ 500 nm )} \\
 \hline
 \hline
         Layer  & Wind direction (angle)   & Wind speed (m/s) & $C_n^2$ & $L_0$(m) \\
 \hline  
 1 & 80.2  & 8.5  & 0.672 & 20 \\
 2 & 90.0  & 6.55 & 0.051 & 20 \\
 3 & 95.7  & 6.6  & 0.028 & 20 \\
 4 & 101.4 & 6.7  & 0.106 & 20 \\
 5 & 177.6 & 22   & 0.08  & 20 \\
 6 & 183.3 & 9.5  & 0.052 & 20 \\
 7 & 189.1 & 5.6  & 0.01  & 20 \\
  \hline  
\end{tabular}
\end{table}

\subsection{Wavefront prediction, noise reduction}\label{sec:pred_and_noise}


The reconstruction for different spatiotemporal priors (for the whole time interval) is obtained with equations described in Section \ref{sec:stgp_posterior}, and the prediction is the marginal distribution of $\phi^{p+2}$. 
The spatial reconstruction (S-GP) is obtained via equations \eqref{eq:sgp_posterior} and \eqref{eq:sgp_mean}. Since this method does not model temporal evolution, the reconstruction does not include the variance from the time delay. Hence, following the standard error budget modeling, we add the temporal variance term to the estimate to obtain the right uncertainty estimate for the reconstruction error. The variance of the change between two frames, that is, the variance of $\phi(x,y,t_1) - \phi(x,y,t_2)$, can be calculated using the spatiotemporal covariance function \eqref{eq:stgp}: 
\begin{align}\label{eq:temporal_variance}
  \sigma_{\rm temp }^2 &= \text{Var}(\phi(x,y,t_1) - \phi(x,y,t_2)) \nonumber \\ &= 2k_\psi(0) - 2 k_\psi((x,y,t_1), (x,y,t_2)).
\end{align}
The variance of the reconstruction error at a given location for this method is then the sum of the appropriate diagonal element in the posterior covariance \eqref{eq:sgp_posterior} and the temporal variance \eqref{eq:temporal_variance}. 

We compare the performance of the FF-GP and WAFF-GP predictive reconstruction to the non-predictive S-GP model by examining the full posterior variance \eqref{eq:stgp_posterior} and the posterior variance filtered with DM influence functions. Since the SH-WFS is a slope sensor that is not sensitive to the piston mode, the uncertainty in the piston mode dominates the posterior variance. Further, the global piston mode does not affect the performance of the AO system. Hence, we filter out the piston mode in the uncertainty and the mean in all comparisons.



\begin{figure}
\centering
	\includegraphics[width=0.9\columnwidth]{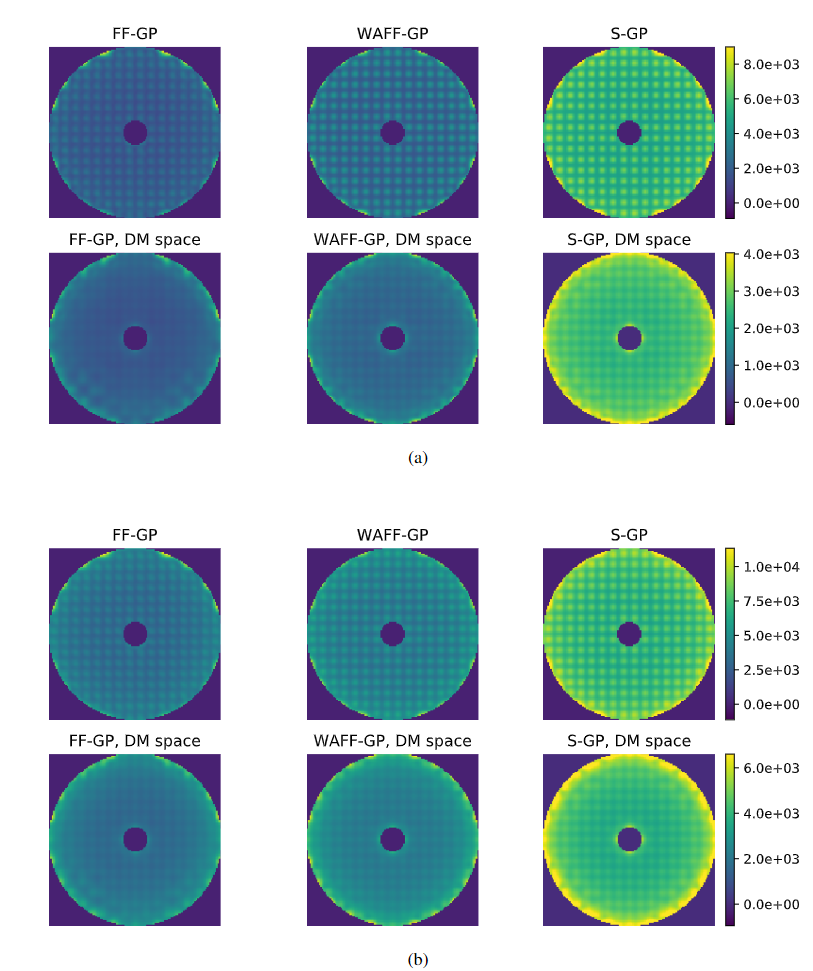}
    \vspace{0.2cm}
    \caption{Piston-free predictive reconstruction accuracy/varinace ($\text{nm}^2$) in low (panel (a), S/N = 100) and high (panel (b), S/N = 4) noise regimes. The upper rows correspond to the full phase estimate, and the lower row images are the least-squares fit to the DM modes. The images illustrate the spatial uncertainty, i.e., the diagonal of the marginal posterior covariance matrix $\phi^{p+2}$, as an image for different methods (FF-GP, WAFF-GP, and S-GP). As S-GP does not predict, the temporal variance of two timesteps has been added to the variance estimate; see \eqref{eq:temporal_variance}.}
    \label{fig:spatial1}
\end{figure}

Figures \ref{fig:spatial1} show the piston-free predictive reconstruction accuracy of low and high noise regimes on phase and DM space, where we discretize the wavefront according to $8 \time 8$ pixels per WFS lenslet. The images illustrate the spatial uncertainty for different methods (FF-GP, WAFF-GP, and S-GP), i.e., the diagonal of the marginal covariance matrix of $\phi_{p+2}$ as an image. We observe that FF-GP delivers the smallest posterior variance, the WAFF-GP the second smallest, and the S-GP the largest for both noise levels, as expected. The posterior variances for WAFF-GP and S-GP are symmetrical since they do not consider the lateral movement in the FF hypothesis. On the other hand, the FF-GP takes into account the advection, and hence, at the edges of the telescope pupil, we see a reduction in variance downstream of the wind. Moreover, the phase reconstruction accuracy (i.e., no DM) of all the methods shows a checker-board-like pattern because the SH operator essentially gives information on the edges of lenslets. The pattern is most pronounced for S-GP and WAFF-GP. The FF can again use the FF advection to lower the uncertainty in the middle of lenslets. 

\begin{figure}
\centering
	\includegraphics[width=0.68\columnwidth]{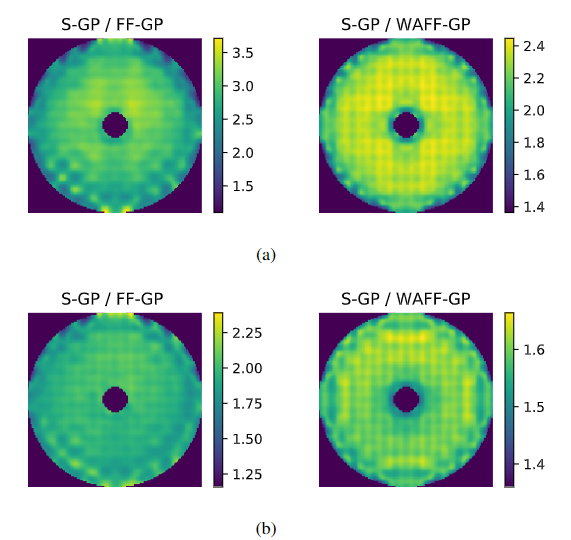}
    \vspace{0.2cm}
    \caption{The relative gain in the variance of the spatial prediction in DM space for the spatiotemporal models in low (panel (a), S/N = 100) and high (panel (b), S/N = 4) noise regimes. The images illustrate the ratio between the baseline spatial prediction (S-GP) variance and the spatiotemporal predictions (FF-GP and WAFF-GP) variances. A significant gain in the posterior variance is observed for both spatiotemporal models.}
    \label{fig:spatial2}
\end{figure}

Figure \ref{fig:spatial2} shows the relative gain of using the predictive reconstruction models compared to the non-predictive S-GP model on DM space. In low noise conditions, FF-GP offers a factor of 3 to 3.5 (depending on the aperture location) improvement in the posterior variance, while WAFF-GP offers a factor of 2 to 2.4 improvement (see Figure \ref{fig:spatial2} (a)). In high noise conditions, the improvement factors are 2.0-2.25 and 1.4-1.8, respectively (see Figure \ref{fig:spatial2} (b)). The performance gain is attributed to three different terms: temporal error, measurement noise, and aliasing. Prediction-wise the MAP estimate gives the minimum variance estimate of the future turbulence. Also, the usage of multiple measurements from the past allows the predictive reconstructor to average the measurement noise. Moreover, the usage of FF advection enables recovering frequencies above the cutoff frequency of the WFS, as we can see from the diminished checkerboard pattern for the FF model in Figure \ref{fig:spatial1}.

\subsection{The effect of modeling errors}
\label{sec:antialias}
The uncertainty estimates derived in the preceding subsection were obtained with the full WFS measurement model accuracy (i.e., $8 \times 8$ pixels inside each lenslet). However, using the full accuracy model comes with a cost in computational complexity in computing the posterior distribution \eqref{eq:stgp_posterior} and the reconstruction matrix \eqref{eq:recontructor}, and deriving such estimates with full accuracy becomes computationally unfeasible when bigger telescopes and longer time series are considered (see Sec. \ref{sec:compelxity}). Here, we examine the effect of the discretization parameter $S$ of the WFS model that defines the discretization grid and, consequently, the accuracy of the model and the computational expense. We consider five different discretization grids: the Fried grid ($S = 1$), an extended Fried grid, where extra pixels are added in between DM actuators ($S = 2 + 1$), $4 \times 4$ pixels inside each lenslet ($S = 4$), $6 \times 6$ pixels inside each lenslet ($S = 6$), and the full $8 \times 8$ grid (see Fig. \ref{fig:fried}). Since our model in \eqref{eq:stgp_posterior} does explicitly account for modeling errors, the accuracy of the WFS model also affects the posterior variance calculations. Hence, instead of uncertainty estimates, we ran a simple Monte Carlo simulation and compared the reconstruction accuracy of the models. The data is created with the full $8 \times 8$ grid ($S = 8$).

\begin{figure}
\centering
	\includegraphics[width=0.9\columnwidth]{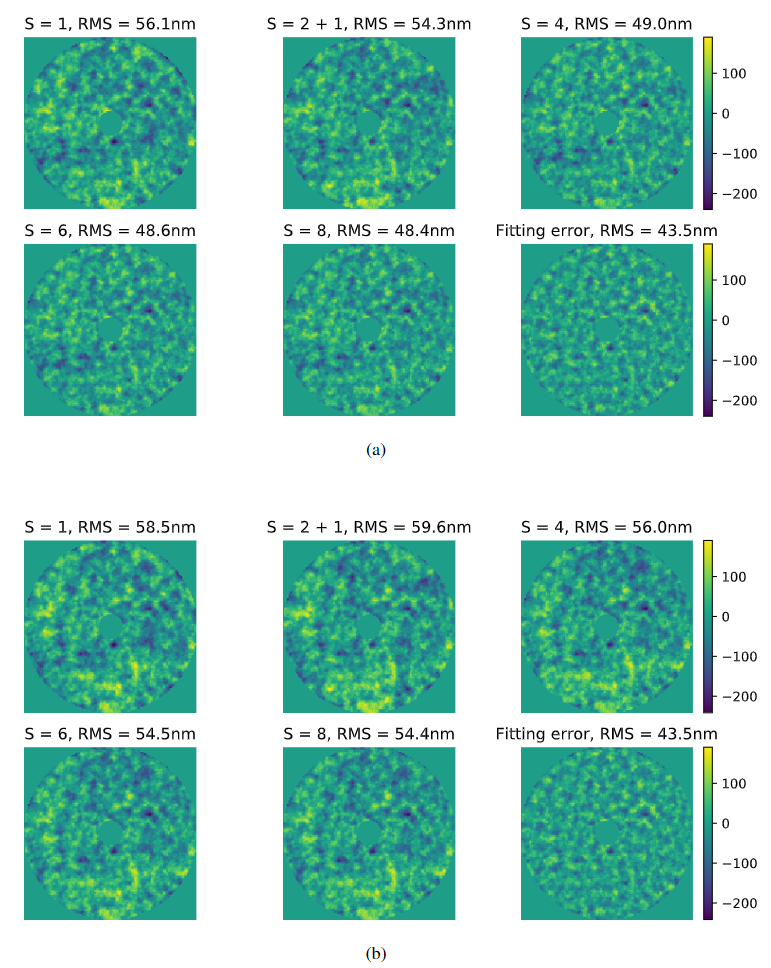}
    \vspace{0.2cm}
    \caption{The effect of the discretization accuracy on the reconstruction accuracy in Monte Carlo simulation. Panel (a): FF-GP prior. Panel (b): WAFF-GP prior.}
    \label{fig:MC}
\end{figure}

Figure~\ref{fig:MC} presents a comparison between the considered discretization levels. Panel (a) is for the FF-prior, and panel (b) is for the WAFF model. The fitting error image in both figures is obtained by projecting $\phi_{p+2}$ to the DM influence functions. As expected, the reconstruction error is small for both models with larger $S$. The WFS model for $S = 1$ basically assumes that a WFS does not measure spatial frequencies above the DM spacing, making it more prone to aliasing error, while the $S =4$ pixels grid provides performance very close to the full discretization grid for both models; Hence all the following experiments are calculated by using $S = 4$. 

\subsection{Utility of measurements}
\label{ssec:oed}

The numerical experiment of this section aims to investigate the utility of including more past timesteps in the prediction for both spatiotemporal models, FF-GP and WAFF-GP. Using a longer time series increases the computational demands of the prediction, and, thus, beyond a certain point, additional data only provides a small benefit; thus, one can limit the number of time steps. In the experiment, the utility of the measurement is computed for time series lengths of $1-16$ for two different noise levels. The utility function used is the square root of the trace of the posterior covariance matrix, i.e., the square root of the Bayesian A-optimality indicator, which measures the expected reconstruction error over the discretization grid projected to DM space. The expected reconstruction errors are compared to the setting where only the prediction's current measurement at $t=0$ is used (e.g., AR model of the first order). The telescope and atmosphere parameters used are the same as in the previous section.

\begin{figure*}\centering
\includegraphics[trim={0cm -0.5cm 0cm 0cm}, width=0.98\textwidth]{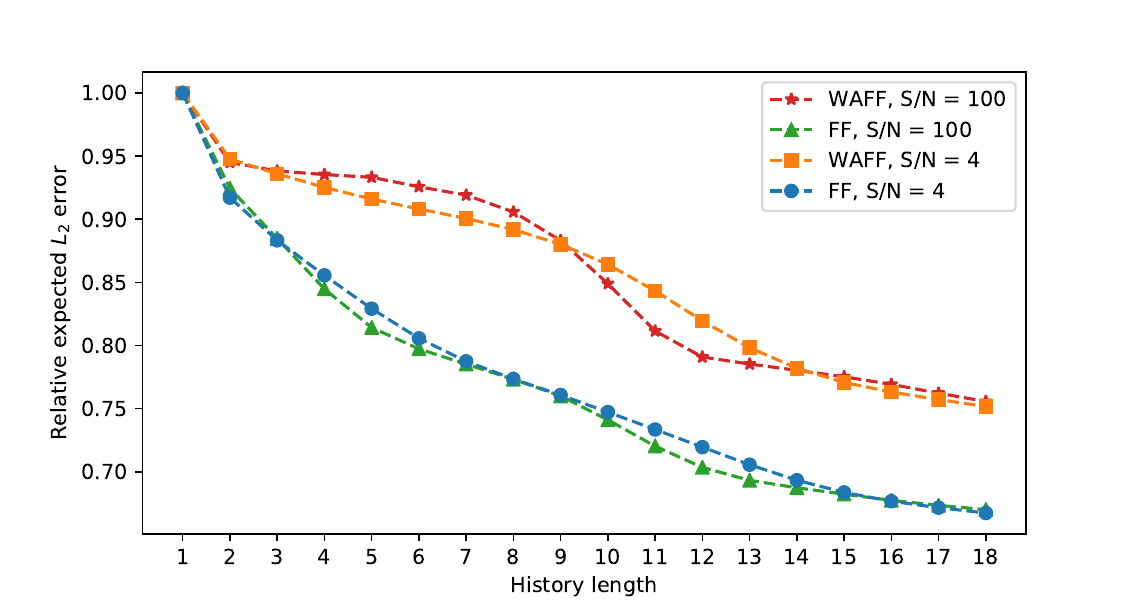}
\caption{Relative expected prediction errors for the FF-GP and WAFF-GP models for different lengths for the time series data in comparison to only using the current state information.}
    \label{fig:oed}
\end{figure*}

We note that since the covariance of the posterior does not depend on the data in the considered Gaussian setting, we do not need to draw data from the prior for this experiment. The results are shown in Figure~\ref{fig:oed}. For the FF-GP model, the quality of the prediction increases more with additional data, and the slope of the curves only starts leveling out near the end of the considered history interval. The improvement is more modest for the WAFF-GP model. 
Be that as it may, considering all data in the studied history range (and beyond) seems advantageous for both prior models as long as it is permitted by the computational constraints.

\subsubsection{Choice of data history}
The limiting factor for using a long history of data is the matrix inversions for deriving the control matrix (Eq. \eqref{eq:stgp_posterior}), not the capability to keep longer time series accessible for the controller. Hence, in addition to the question on a useful history length, another important question is: which previous time steps should be used in the prediction? As the considered Gaussian priors exhibit structures of a certain size, it may not be optimal (given limitations in computational resources) to use only the most recent available data but to rather choose more sparse temporal presentations from the past data.

To investigate this, we compute an optimal combination of history data to be included in the prediction by resorting to a greedy algorithm that iteratively includes data from past timesteps in order to minimize the associated expected reconstruction error. The posterior covariance matrix is formed and projected to the DM space, the A-optimality target is computed for each possible addition, and the choice that yields the lowest target value is added to the timeseries. The algorithm can be iteratively continued to include data from as many past timesteps as desired.
In the numerical experiment, we compute the first five optimal choices; to be precise, the first included data always corresponds to the latest timestep, $t = 0$, and the other four timesteps are chosen with the algorithm. We note that choosing the data in a greedy manner does not guarantee that the choices would be globally optimal, but such an optimization approach is adopted due to computational limitations.

\begin{figure*}
\centering
\includegraphics[trim={0cm -0.5cm 0cm 0cm}, width=0.97\textwidth]{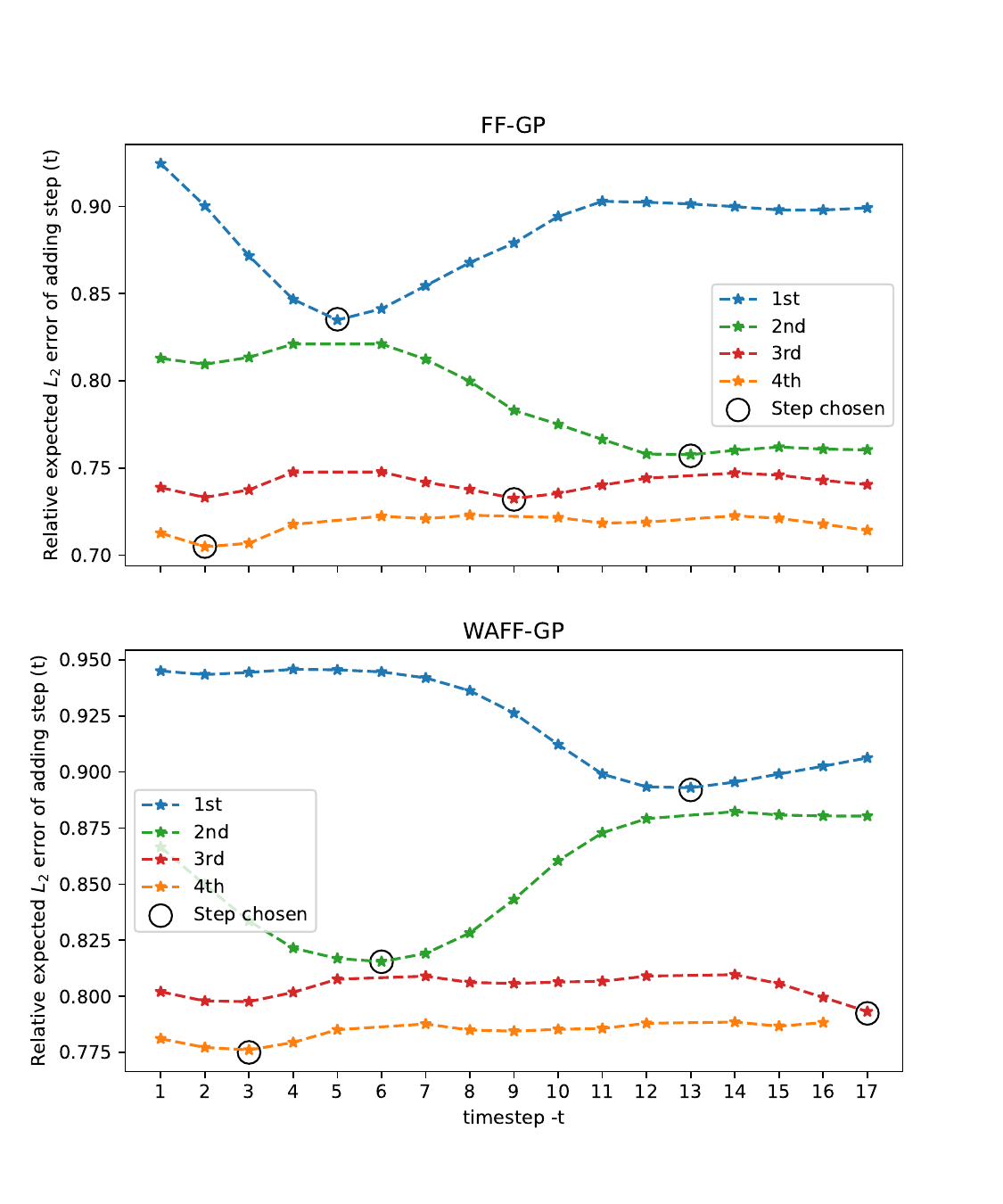}
\vspace{0.2cm}
\caption{The optimization process for choosing the most informative timesteps from the past. The upper plot is for the FF model, and the lower plot is for the WAFF model (framerate 2ms and S/N = 100). At every optimization step (1st, 2nd, 3rd, and 4th), we plot the expected $L_2$ error of adding the step (t) compared to the expected  $L_2$ error of just including step 0 (relative to first-order prediction). The black circles indicate the chosen step, which is then omitted in subsequent calculations.}
\label{fig:oed2}
\end{figure*}

Figure \ref{fig:oed2} shows the target values for the first four iterations of the greedy algorithm for both the FF-GP and the WAFF-GP models, with 2ms framerate and a signal-to-noise ratio of 100. As expected, choosing the latest data is not the optimal solution; instead, it seems to be better to choose data that is temporally sparser than the framerate at which the system operates. 

The chosen timesteps, in the order that they were included by the algorithm, is $[0,  5, 13,  9,  2]$ for the FF-GP and $[0, 13,  6, 17,  3]$ for the WAFF-GP. In both cases, the final sampling of the data corresponds to intervals of roughly $6-8$ms (2-4frames). The optimal timesteps for the FF-GP lead to a relative expected error of 0.725, which corresponds to the same value as using the 11--12 latest consecutive timesteps and gives a 10\% improvement compared to using just five latest consecutive steps; cf.~Figure~\ref{fig:oed}. For the WAFF model, the corresponding numbers are 13--14 steps and 15\%.

We also experimented with how the framerate and measurement noise affect the optimal use of history data. A faster framerate favors sparser presentation while increasing the measurement noise leads to more dense temporal sampling. 

\section{Discussion}
We present a predictive approach based on GP modeling. In the context of the FF assumption and linear wavefront sensing, the presented FF-GP model is optimal in the least-squares sense conditioned on the fixed times series of WFS data and the specified spatiotemporal 
(Von K{\`a}rm{\`a}n) prior to the turbulence. Hence, the derived improvements correspond to the limits achievable by predictive control for Von  K{\`a}rm{\`a}n turbulence. We also studied a less informative model, WAFF-GP, with only coarse assumptions about the atmosphere. These models allow a closed-form estimate of how good reconstruction and prediction can ideally be achieved with given assumptions on the telescope geometry and atmospheric conditions and our knowledge of them. 


As mentioned above, the linear predictive filter/reconstructor $R_{\rm FF}$ is optimal in the least-squares sense conditioned on the fixed time series of WFS data and the specified spatiotemporal prior for the turbulence. Consequently, the results show that non-linearity is only encountered in predictive control under the FF hypothesis when non-linear WFSs are considered. 

Further, the results indicate that utilizing spatiotemporal correlations increases the prediction accuracy in numerical simulations, reducing variance up to a factor of 3.5 compared to a non-predictive approach, while an uncertain wind profile leads to an improvement of 2.3, which aligns well with the theoretical limits in \citenum{doelman2020minimum} and on-sky results on Keck II\cite{van2022predictive}. However, as discussed in \cite{van2020robustness}, it is unclear if these spatiotemporal correlations provide performance gain on real-world data. The predicted performance gain depends on various aspects, such as atmospheric conditions, WFS used, the WFS model accuracy, telescope geometry, and sampling rate. Moreover, if the predictive controller (e.g., EOF) is learned from simulated data, the predicted performance gain depends on the way the data is simulated. For example, the temporal interpolation of the FF turbulence screens dampens higher frequencies from the temporal spectral, leading to smoother, more predictable turbulence. Also, disentangling aliasing error and temporal error by using an idealized phase sensor can lead to more optimistic performance gains.

The numerical simulations were conducted with a fairly small  $16 \times 16$ system. However, since the turbulence is spatially isotropic and the WFS model operates locally, the conclusions translate (approximately) to bigger systems with similar actuator spacing, e.g., SPHERE.

Moreover, our results indicate it is always --- or at least inside the preceding 18 frames and with the tested model parameters --- advantageous to include more history steps into the model if permitted by the available computational resources. However, at around 16 frames, the gain from including extra timesteps starts to level out slowly. Moreover, we demonstrated that given maximum history length, a sparse sampling of past data over a longer time span leads to better predictions than a sequence of the last consecutive frames. An optimized choice of five history steps provides in the studied setting an additional 10-15\% improvement in the RMS, indicating that optimized use of history data may also be important for machine learning-based minimum variance predictive control, such as empirical orthogonal functions (EOF) \cite{guyon2017adaptive}.

 

\subsection{Computational complexity}
\subsubsection{Non-real-time computations: computing the posterior distribution}\label{sec:compelxity}

To apply the predictive reconstruction matrix \eqref{eq:recontructor}, one must solve for (or be able to operate with) the posterior covariance matrix \eqref{eq:stgp_posterior}. The calculation of the inverse of the dense matrix in \eqref{eq:stgp_posterior}is computationally expensive. For a dense symmetric positive definite matrix of size $n \times n$, a standard matrix inversion method, say, the Cholesky decomposition, has a computational complexity of $\mathcal{O}(n^3)$. Hence, operating with the inverse of a matrix of an 8-meter class telescope already requires substantial memory resources, and the computational requirement would be in the order of teraFLOPs or more. As an example, 8-meter-class XAO ($40 \times 40$ DM) systems with $p = 5$ past telemetry require inverting a matrix of $91524 \times 91534$ when $S = 4$ and  $7752 \times 7752$ when the $S = 1$ (Fried grid), if the size of the matrix is optimized without any approximations. Performing such computations efficiently might involve specialized hardware, parallel processing, and optimized algorithms to handle the scale of the problem efficiently, especially when DoF is scaled up to $10^4$, in the next generation of extremely large telescopes. 

Since the posterior variance does not depend on the data, the predictive reconstructor \eqref{eq:recontructor} needs only be updated when the prior information (e.g., the wind direction, wind speed, or $r_0$) changes. The presumed rate of change in the atmospheric parameters determines the update rate for the posterior covariance.


A connection exists between the method discussed here and EOF, which uses a batch of past data to fit the spatiotemporal covariates of pseudo-open-loop telemetry. If we replace the WFS model with an ideal phase sensor (identity matrix), the EOF prediction matrix will converge to the prediction matrix derived from the FF model on the infinite data limit (stable atmospheric statistics). The EOF is machine learning based, so it can potentially adapt to more general turbulence statistics, while ST-GP formulation proved a way to deal with aliasing error and study the theoretical limits of predictive control.



\subsubsection{Real-time computations: applying the mean prediction matrix}
The real-time computation needed for AO control is the mean prediction of the marginal distribution corresponding $\bm \phi^{p+2}$, i.e., a matrix multiplication between the past sequence of WFS data and the predictiton matrix $R_{\rm FF/WAFF}$ defined in \eqref{eq:recontructor}. The reconstruction matrix shape is the $1292 \times (p \times 2400) = 1292 \times 12000$, if $p=5$. Again, the prediction matrix need not be recomputed as long as the atmospheric conditions stay the same. Assuming dense matrices, multiplication with the prediction matrix requires a few hundred Gflops of computing bandwidth, which falls comfortably within the capabilities of conventional computers available today.


\subsection{Conclusion and future work}
To conclude, this paper studies the limits of predictive control with spatiotemporal Gaussian process models. It discusses how reconstruction errors, such as temporal error, photon noise, and aliasing, can be minimized with predictive control in an optimized manner, given the computational limitation of the hardware used. It also studies how modeling errors, particularly WFS model discretization, affect the quality of the predictive controllers. 

The concepts presented in this paper offer several avenues for future research. This work assumes that the knowledge of atmospheric parameters (either the full wind profile or just $r_0$ and $\tau_0$ ) is given a prior. However, it would also be possible to incorporate their estimation into the algorithm by modeling them as random parameters and computing a \textit{maximum likelihood} estimate at each measurement step. This is a considerable advantage of the Bayesian approach: including additional sources of uncertainty can be done systematically. An especially interesting direction would be to see if we can recover FF parameters from on-sky data (e.g., SPHERE telemetry) and utilize them in predictive control.

Also, the application of experimental design to AO could be investigated further, and many other ways of optimizing the measurement could be considered. An interesting approach could be to discretize the data inhomogeneously along the spatial and temporal axes depending on its informativeness. Since older data can be assumed to be less important, one could consider coarsening the discretization for older timesteps as new data is introduced. Moreover, our approach provides a systematic way to include stochastic vibrations in the model and to use Bayesian OED to ensure that the data relevant for predicting the vibrations is included in the prediction process. 

This paper considers a simplified AO design with an SHS sensor operating in an open-loop setup. Theoretically, the method adapts to a closed-loop system via the pseudo-open-loop scheme, which is already provided by some real-time systems working on-sky. However, pseudo-open-loop adaption requires knowledge of hardware/software time lags, deformable mirrors' and wavefront sensors' response times, as well as calibration errors. Any bias in these reduces the method's performance and can lead to instabilities in the closed-loop control. Although this paper does not explore the effect of the mentioned error sources, it is important to consider them when implementing and deploying the technique on real hardware. 

Finally, these concepts can be used to study speckle statistics under predictive control. The predictive control uncertainties could be propagated through a coronagraphic system to give the WDH intensity under optimal predictive control with the given assumption of the atmosphere. Overall, the presented method not only enhances the accuracy and resolution of reconstructions but also opens new avenues for advancing predictive control and reconstruction methodologies and provides a deeper understanding of the limits of given prediction models, e.g., those based on machine learning.
 






\section*{Acknowledgements}
The work of JN and TH was supported by the Academy of Finland (decisions 326961, 345720, and 353094). The work of NH and JP was supported by the Academy of Finland (decisions 348503,353081, 359181). We thank Miska Le Louarn for fruitful discussions on the project.

\section*{Data Availability}
The data and codes used in this paper are available on GitHub in Jupyter Notebook format. The data is completely numerically simulated and produced by the codes. Please visit our GitHub repository  [\href{https://github.com/jnousi/ST-GP4AO.git}{https://github.com/jnousi/ST-GP4AO.git}] to access the codes. The repository contains Jupyter Notebook files that outline the analysis steps taken in this study. The code is documented and annotated to help readers understand the methodology and reproduce the results.

We encourage readers to use the data and codes for their own research and to cite this paper as the source of the data. If you have any questions about the data or the codes, please do not hesitate to contact us.

\bibliography{report}   
\bibliographystyle{spiejour}   


\vspace{2ex}\noindent\textbf{Jalo Nousiainen} is a postdoctoral researcher at Aalto University. His research focuses on algorithm development for high-contrast instruments dedicated to exoplanet imaging. He has a background in applied mathematics, specifically in Inverse problems, Bayesian statistics, and machine learning.

\vspace{2ex}\noindent\textbf{Juha-Pekka Puska } is a doctoral researcher at Aalto University. His research is on optimal experimental design methods and their applications to imaging problems. 

\vspace{1ex}
\noindent Biographies and photographs of the other authors are not available.

\listoffigures
\listoftables

\end{spacing}
\end{document}